\documentstyle[12pt,epsfig]{article}

\begin{document}

\baselineskip=6.8mm

\newcommand{\TeV}{\,{\rm TeV}}
\newcommand{\GeV}{\,{\rm GeV}}
\newcommand{\MeV}{\,{\rm MeV}}
\newcommand{\keV}{\,{\rm keV}}
\newcommand{\eV}{\,{\rm eV}}
\newcommand{\Tr}{{\rm Tr}\!}
\renewcommand{\arraystretch}{1.2}
\newcommand{\be}{\begin{equation}}
\newcommand{\ee}{\end{equation}}
\newcommand{\bea}{\begin{eqnarray}}
\newcommand{\eea}{\end{eqnarray}}
\newcommand{\ba}{\begin{array}}
\newcommand{\ea}{\end{array}}
\newcommand{\bmat}{\left(\ba}
\newcommand{\emat}{\ea\right)}
\newcommand{\refs}[1]{(\ref{#1})}
\newcommand{\ler}{\stackrel{\scriptstyle <}{\scriptstyle\sim}}
\newcommand{\ger}{\stackrel{\scriptstyle >}{\scriptstyle\sim}}
\newcommand{\lag}{\langle}
\newcommand{\rag}{\rangle}
\newcommand{\ns}{\normalsize}
\newcommand{\cm}{{\cal M}}
\newcommand{\gr}{m_{3/2}}
\newcommand{\p}{\partial}
\def\tl{{\tilde{l}}}
\def\tL{{\tilde{L}}}
\def\bd{{\overline{d}}}
\def\tL{{\tilde{L}}}
\def\a{\alpha}
\def\b{\beta}
\def\g{\gamma}
\def\c{\chi}
\def\d{\delta}
\def\D{\Delta}
\def\db{{\overline{\delta}}}
\def\Db{{\overline{\Delta}}}
\def\e{\epsilon}
\def\l{\lambda}
\def\n{\nu}
\def\m{\mu}
\def\nt{{\tilde{\nu}}}
\def\p{\phi}
\def\P{\Phi}
\def\x{\xi}
\def\r{\rho}
\def\s{\sigma}
\def\t{\tau}
\def\th{\theta}
\renewcommand{\Huge}{\Large}
\renewcommand{\LARGE}{\Large}
\renewcommand{\Large}{\large}

\rightline{IC/97/113 }

\rightline{hep-ph/9708403}


\begin{center}
{\Large \bf Vacuum oscillations of solar neutrinos: 
correlation between spectrum distortion and seasonal variations} 

\vspace{2cm}

{S. P. Mikheyev$^{* , \#}$  and           
A.~Yu.~Smirnov$^{\dagger,\#}$ \\[.5cm]
 
{\ns\it $^*$ Laboratory Nazionale del Gran Sasso dell'INFN}\\
{\ns\it I-67010 Assergi (L'Aquilla), Italy} \\[.3cm]

{\ns\it $^\dagger$ International Center for Theoretical Physics}\\
{\ns\it P.~O.~Box 586, 34100 Trieste, Italy} \\[.3cm]

{\ns\it $^\#$ Institute for Nuclear Research, Russian Academy of Sciences}\\
{\ns\it 117312 Moscow, Russia} }
\vspace{2cm}

\begin{abstract} \baselineskip=7mm

{\ns 

Long length  vacuum oscillations solution 
 of the solar neutrino problem  
is discussed. We show that  there is a strict correlation between 
a distortion of the neutrino energy spectrum and an amplitude 
of seasonal variations of the neutrino flux.  
The {\it slope} parameter which characterizes a distortion of the
recoil electron energy spectrum in the Super-Kamiokande experiment 
and the seasonal asymmetry of the signal  have been calculated 
in a wide range of  oscillation parameters. 
The correlation of the slope and asymmetry  gives crucial 
criteria for identification or exclusion 
of  this solution.  For the positive slope indicated by 
preliminary Super-Kamiokande data we predict  (40 - 60) \% enhancement 
of the seasonal variations. 
}
\end{abstract}

\end{center}

\thispagestyle{empty}

\newpage

1. Long length vacuum oscillations of neutrinos on the way from
the Sun to the Earth are considered as a viable solution of 
the solar neutrino problem  \cite{pon,grib,pom,barg,glash} 
(see \cite{kras,calab,zur,hata} for latest analysis) 
\footnote{Large mixing angles implied by 
this solution are  however disfavored by the data from the SN87A
\cite{ssb}.  }. 
There are two key signatures of this  solution:     
(i) distortion of the neutrino energy spectrum 
\cite{barg,glash}, and  
(ii) seasonal variations of the fluxes \cite{pom}.    

(i). The distortion follows from a dependence of the oscillation 
survival probability, $P$, on the neutrino energy $E$:  
\be
P = 1 - \sin^2 2\theta \sin^2 \left( \frac{\Delta m^2}{4}
\frac{L}{E} \right)~.
\label{prob}
\ee
(We consider mixing of two  neutrinos.) 
Here $\Delta m^2 \equiv m_2^2 - m_1^2$ is 
the neutrino mass squared difference,  
$\theta$ is the  mixing angle 
and $L$ is the distance between the Sun and the Earth.     
A variety of distortions 
of the boron neutrino spectrum 
is expected depending on values of 
$\Delta m^2$ and $\sin^2 2\theta$.

(ii). Seasonal variations are stipulated by ellipticity of the
Earth's orbit. 
The flux of neutrinos at the Earth, $F$,  can be written as 
\be
F = F_0 \left( \frac{L_0}{L} \right)^2 \cdot P(L, E), 
\label{flux}
\ee
where $F_0$ is the flux at  the astronomical unit $L_0$.  
The distance between the Sun and the Earth     
at a given moment $t$ equals: 
\be
L \approx L_0 \left( 1 + \epsilon \cos \frac{2\pi t}{T} \right)^{-1} ,  
\label{dist}
\ee
$\epsilon = 0.0167$ is the 
eccentricity, $T \equiv 1$ year.  
The variations are expected both due to  the  
geometrical factor, $1/L^2$, and due to change of 
the oscillation probability  
\cite{pom,glash}. Depending on 
values of the oscillation parameters 
$\Delta m^2$ and  
$\sin^2 2\theta$ one may get an  enhancement or 
damping of the geometrical  effect \cite{pet,faid,fogli} or  
even more complicated variations of signals.

In this Letter we point out that there is a strict 
correlation between a distortion 
of the boron neutrino energy spectrum and seasonal variations  
of  the  boron neutrino flux. This gives crucial 
criteria for identification or discrimination of  
vacuum oscillations solution.\\

2. The correlation between the distortion and time variations 
can be immediately seen from the expression for the oscillation 
probability. Indeed, Eq. (\ref{prob}) gives  
\be
\frac{d P}{d L} = - \frac{d P}{d E} \cdot \frac{E}{L}~
= - s_{\nu} \cdot \frac{P E}{L},     
\label{relation}
\ee 
where 
\be
s_{\nu}(E) \equiv \frac{1}{P}\frac{dP}{dE}.   
\ee 
is the slope parameter   
which  characterizes the distortion of the neutrino spectrum at 
a given energy $E$.  
If  spectrum distortion has a positive slope, 
$s_{\nu} > 0$, 
then according to Eq. (\ref{relation}) one gets    
${d P}/{d L} < 0$, {\it i.e.} with increase of the distance the 
survival probability   decreases. 
The  change of the flux is in the same direction as due to 
the geometrical factor. Therefore,  for a positive 
slope the vacuum oscillations  enhance  seasonal 
variations. For a negative slope,  
$s_{\nu} < 0$, 
Eq. (\ref{relation}) gives 
${d P}/{d L} > 0$, that is,  
the probability increases with distance. In this case 
the oscillations 
weaken   seasonal variations due to pure  geometrical factor.

Let us find the correlation explicitly. 
From  Eqs. 
(\ref{prob},\ref{flux}) 
we get  
the   change of  neutrino  flux with distance:  
\be
\frac{d F}{d L} = - \frac{F}{L} \left( 2 + 
\frac{d P}{d E} \cdot \frac{ E }{ P }   
\right).   
\label{dflux}
\ee
Here the first term is due to the geometrical 
effect and the second one corresponds to a 
change of the probability with distance. 
For a positive slope both terms  have the same sign 
in accordance with previous discussion.    
Introducing  the seasonal asymmetry: 
\be
A_{\nu} = 2\frac{\Delta F}{\bar F}, 
\label{asym}
\ee 
where $\Delta F$ is the difference of the averaged fluxes 
during the winter and the summer and $\bar F$ is the averaged 
flux during the year,  we can define the quantity 
\be
r_{\nu} \equiv \frac{A_{\nu} - A_{\nu}^0}{A_{\nu}^0}, 
\label{rasym}
\ee
with  
$A^0_{\nu} \equiv A_{\nu}(P = 1)$ being the asymmetry without
oscillations. The asymmetry $r_{\nu}$ gives the variations due to
the oscillations in the
units of pure geometrical effect. 
From (\ref{dflux}) we get final relation between   
the  slope parameter and the oscillation asymmetry $r_{\nu}$: 
\be
s_{\nu} = \frac{2}{E} r_{\nu} . 
\label{corr}
\ee 
Let us stress that this 
relation does not depend on oscillation parameters 
$\Delta m^2, \sin^2 2\theta$ at least in 
the lowest order on the eccentricity $\epsilon$. \\

3. The  correlation  described by  (\ref{dflux}, \ref{corr}) 
holds for fixed neutrino energy. 
In real experiments the energy spectrum  of the recoil (produced) 
electrons is measured, and moreover, the integration over 
certain energy intervals of neutrinos as well as electrons 
(due to finite energy resolution) takes place. 
This  modifies the correlations.  

Let us consider a  manifestation of the correlation 
in the Super-Kamiokande experiment \cite{SK,SK3}.  
For a majority of relevant values of 
$\Delta m^2$ and $\sin^2 2\theta$ the  
whole observable part of the boron neutrino spectrum 
(6 - 14 MeV) is  
in  regions  with definite sign of ${d P}/{d E}$. 
Integrations over the neutrino energy and on the electron energy 
weighted by the energy resolution function lead to 
a strong smoothing  of the observable  distortion of the recoil electron
spectrum. 
The  distortion can be well characterized by  a sole  
{\it slope parameter}  $s_e$  \cite{rosen}   defined as:  
\be
\frac{N_{osc}}{N_0} \approx R_0  + s_e T_e  , 
\label{slope}
\ee  
where $N_{osc}$ and $N_0$ are the numbers of events with 
and without oscillations correspondingly, 
$R_0$ is a constant,  
$T_e$ is the recoil electron energy in MeV,   
$s_e$ is in the units  MeV$^{-1}$.  There is some  change of
the spectrum distortion  during the year due to variations of distance
$L$ \cite{pet}. However, this effect is small and can be neglected in the 
first approximation. 

The slope parameter is determined in the
following way: 

(i) We calculate expected numbers of events, $N_{osc}^i$,  
for different values  
$\sin^2 2\theta$ and $\Delta m^2$ in 
the energy  bins  with $\Delta T_e = 0.5$ MeV  
from 6.5 MeV to 15 MeV 
(as in the Super-Kamiokande presentation of the data): 
\be
  N_{osc}^i(L) = 
  \left( \frac{L_0}{L} \right)^2 \int_{T^i_{e}}^{T^i_{e} + \Delta T_e} 
dT_e     \int dT'_e 
  f(T_e^i, T'_e) \int_{T'_e} d E \frac{d \sigma (E, T_e')}{d T_e'} F (E) 
  P(E,L, \Delta m^2, \theta) ,
\label{int}
\ee
where $f(T_e, T'_e)$ is the energy resolution function, 
$d\sigma (E, T_e')/dT_e'$ is the differential cross-section of the 
$\nu e- $ scattering.

(ii) Similar numbers of events,   
$N_0^i$,   have been calculated in absence of 
oscillations: $N_0^i = N^i(P = 1)$. 
Then ratios $N_{osc}^i/N_0^i$ 
have been found for each bin. 

(iii) The $\chi^2$ fit of the histogram $N_{osc}^i/N_0^i$ 
by the function (\ref{slope})    
gives $s_e$ (fig. 1 a). \\

We will characterize the seasonal variations by 
the summer-winter asymmetry: 
\be
A_e \equiv 2 \frac{N_{W} - N_{S}}{N_{SP} + N_{A}} .  
\label{assym}
\ee 
Here  $N_{W}$,  $N_{S}$ , $N_{SP}$, $N_{A}$ 
are the  numbers of events detected
from November 20 to February 19, from May 22 to August 20,
February 20 to May 21, from August 21 to November 19
respectively:  
\be
N_V = \int_V dt \sum_i N_{osc}^i (L(t)) ~~~ (V \equiv W, S, SP, A).   
\label{num}
\ee
Similar asymmetry is calculated in absence of oscillations,  
{\it i.e} due to  pure geometrical factor: 
$A_e^0 =  A_e (P = 1)$.   

Finally,  we introduce  
(as in (\ref{rasym})) the  signal  
asymmetry 
\be
r_e \equiv \frac{A_e}{A_e^0} -1,  ~~~~ 
\ee
which characterizes the asymmetry 
due to oscillations  in the units of geometrical 
asymmetry. Positive $r_e$ corresponds to an enhancement of 
the geometrical effect.\\  

4. The correlation between the spectrum distortion and seasonal variations 
appears as the correlation between the slope parameter $s_e$  and
the signal asymmetry  $r_e$ (fig. 1 - 4). 

In fig.1 we show the  dependence of  the slope and  asymmetry 
on $\Delta m^2$ for different values of 
$\sin^2 2\theta$. Both the slope and  the asymmetry increase 
with  mixing angle, whereas  
 positions of maxima and zeros (in $\Delta m^2$ scale)  do not depend on 
$\theta$. Note that zeros of the asymmetry are shifted 
with respect to zeros of the slope 
by approximately 8\% towards bigger values of $\Delta m^2$. 
Maximal positive asymmetry is also shifted towards bigger 
$\Delta m^2$ (about 25\% for the first maximum), however 
this difference diminishes for next maxima. 
In contrast, first negative maximum of 
asymmetry is shifted towards smaller 
$\Delta m^2$. These features are related to the integration over 
the neutrino and electron energies. 
Maximal asymmetry due to oscillations can 
exceed the geometrical asymmetry ($|r_e| > 1$).

Fig. 2 shows the slope - asymmetry plot. The points correspond 
to different values of $\Delta m^2$ 
and $\sin^2 2\theta$.  Scattering of the points 
is related to the integration over 
the interval of the neutrino energies. 
According to fig.2 the  correlation can be approximated by  
\be
s_e   = (0.03 - 0.10)~{\rm MeV}^{-1}~ r_e~. 
\ee
Note that for the neutrino spectrum  the slope is 
larger: according to (\ref{corr}) 
$s_{\nu} = 0.2~ {\rm MeV}^{-1} r_{\nu}$ 
($r_{\nu} \sim r_e$) for $E = 10$ MeV.

Iso-asymmetry (a) and iso-slope (b) lines in  
$\Delta m^2$, $\sin^2 2\theta$ 
plot are shown in fig.3.  The iso-asymmetry plot 
(fig. 3a) is similar to the iso-plot of the near-far asymmetry 
found in \cite{faid}.\\

Preliminary Super-Kamiokande data for 306 days \cite{SK3} 
show a {\it  positive slope}:    
$ s_e \sim  (0.013 - 0.017)~ {\rm MeV}^{-1}$.  
According to figs. 1, 3, 
in the region  of the best fit of the integral 
data ($\Delta m^2 = (5 - 8) \cdot 10^{-11}$ eV$^2$)   
one expects for these  values of the slope the  
positive asymmetry $r_e = 0.4 -  0.6$ .  
That is,   an {\it enhancement}  of the 
seasonal variations by (40 - 60) \% should be observed. 

In fig. 4 we show real time seasonal variations of the flux 
for different values of the neutrino parameters. 
Dashed-dotted line corresponds to very weak 
variations due to oscillation effect 
and therefore practically coincides with geometrical effect. 
Two other lines correspond to enhancement and suppression 
of the geometrical effect by oscillations. Solid line 
shows  variations  expected from the observed slope.    
Also shown are the preliminary results from the Super-Kamiokande 
(306 days) experiment \cite{SK3} . 
Obviously, present statistics does not
allow one to make any conclusion. 
Significant result can be obtained after 4 - 5 years  
of the detector operation. 

In larger interval of the recoil electron  energies (from  5 MeV to 15 MeV)    
which will be accessible soon 
a description of the distortions by only one parameter may not be 
precise. Moreover, seasonal variations differ 
at different  energies. In this case 
one can divide the interval into 
two parts: {\it e.g.},  (5 - 8) MeV and (8 - 15) MeV and study the  
correlations in these  intervals  separately. 
A comparison of the effects will give the cross-check of the results.

Similar correlation picture can be obtained for SNO experiment 
\cite{SNO} where the effect is expected to be even more 
profound due to weaker averaging.

\bigskip

\noindent
{\bf Acknowledgment:} 

Authors are grateful to E. Kh. Akhmedov and E. Kearns for discussion.

{\begin{figure}[htb]
\center
{\psfig{figure=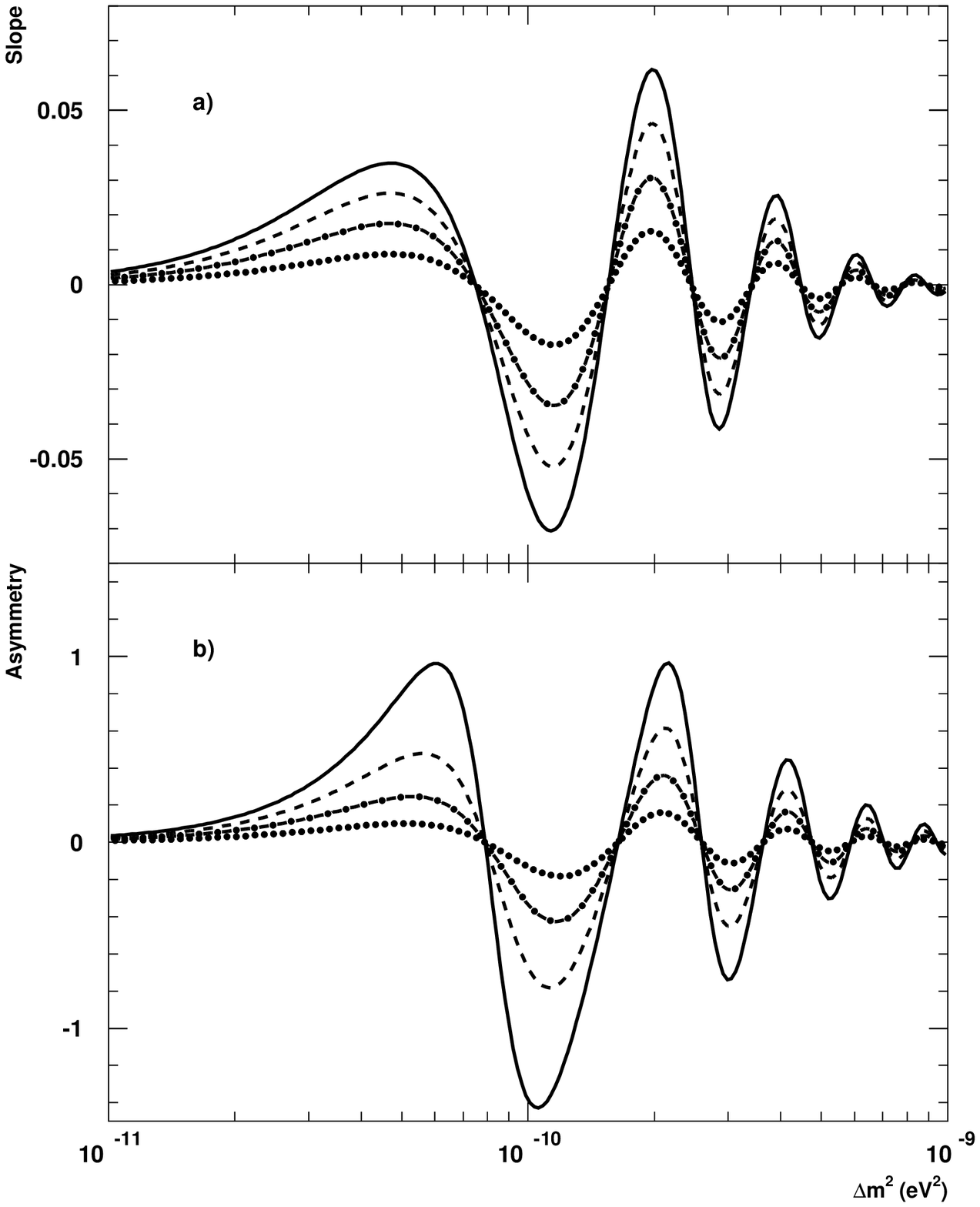,height=20.0cm}}
\protect\caption[]
{
The dependence of (a) the slope parameter $s_e$ (in MeV $^{-1}$) 
 and (b) the signal asymmetry $r_e$  
on $\Delta m^2$ for different values of $\sin^2 2\theta$.
Solid , dashed, dash-dotted, and dotted lines correspond to 
$\sin^2 2\theta$ = 1.0, 0.75, 0.5, and 0.25 respectively. 
}
\end{figure}}

{\begin{figure}[htb]
\center
{\psfig{figure=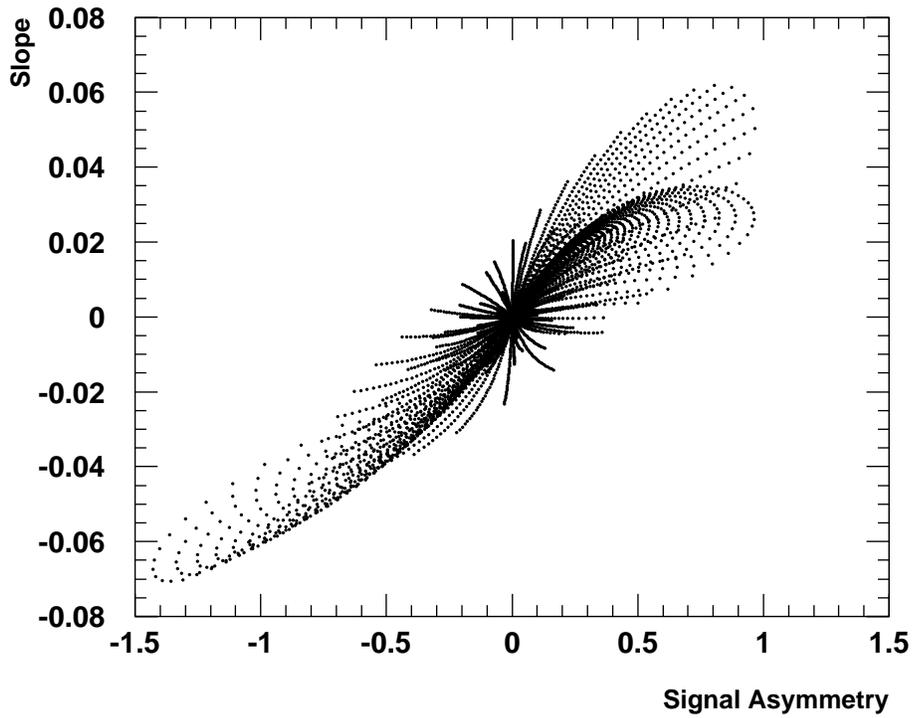,height=10.0cm}}
\protect\caption[]
{
The slope -  asymmetry plot. The points ($s_e - r_e$) correspond 
to different values of $\Delta m^2$ 
(200 points between $10^{-11}$  and   $10^{-9}$ eV$^2$),   
and $\sin^2 2\theta$ (40 points between 0.025 and  1.0). 
Calculations have been done on grid $200\times 40$.  
}
\end{figure}}

{\begin{figure}[htb]
\center
{\psfig{figure=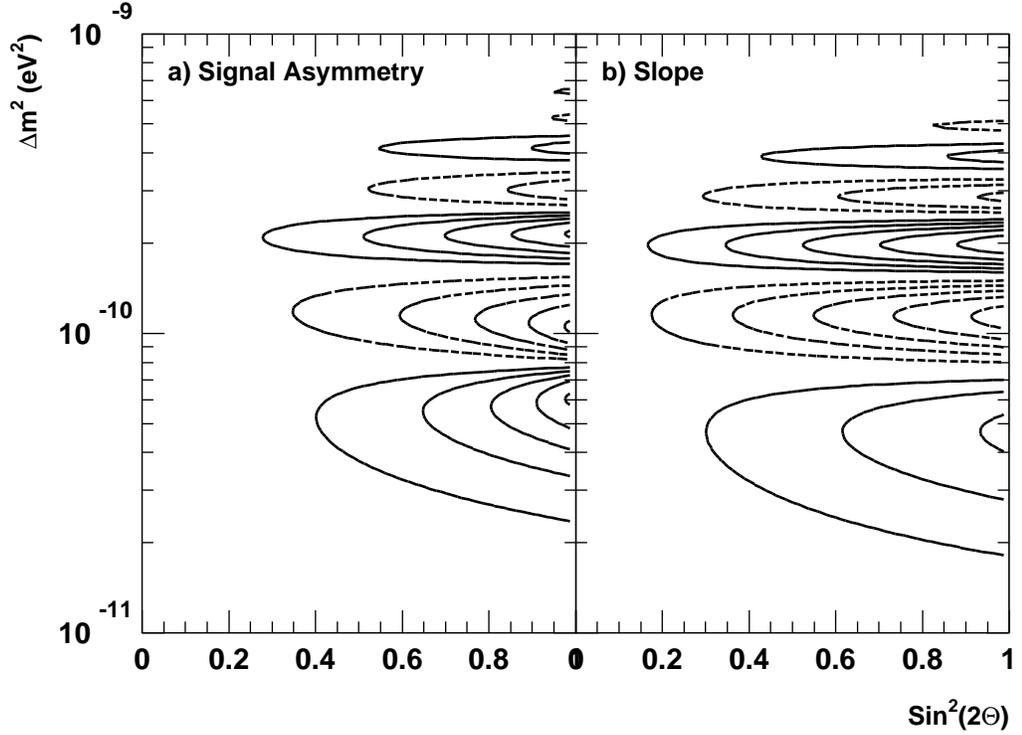,height=10.0cm}}
\protect\caption[]
{
a). Iso-asymmetry lines in $\Delta m^2$, $\sin^2 2\theta$ plot. Solid
lines 
correspond to positive asymmetries: $r_e$ = 0.19, 0.38, 0.57, 0.76, 0.95
(from outer to inner lines). 
Dashed lines correspond to negative asymmetries: $- r_e$ = 0.28, 0.56,
0.84, 1.12, 1.4.

b). Iso-slope  lines in $\Delta m^2$, $\sin^2 2\theta$ plot. Solid lines 
correspond to positive values of the slope parameter: 
$s_e$ = 0.011, 0.022, 0.033, 0.044, 0.055 (from the outer to inner lines). 
Dashed lines correspond to negative slopes: $- s_e$ = 0.013, 0.026,
 0.039, 0.052,  0.065. }
\end{figure}} 

{\begin{figure}[htb]
\center
{\psfig{figure=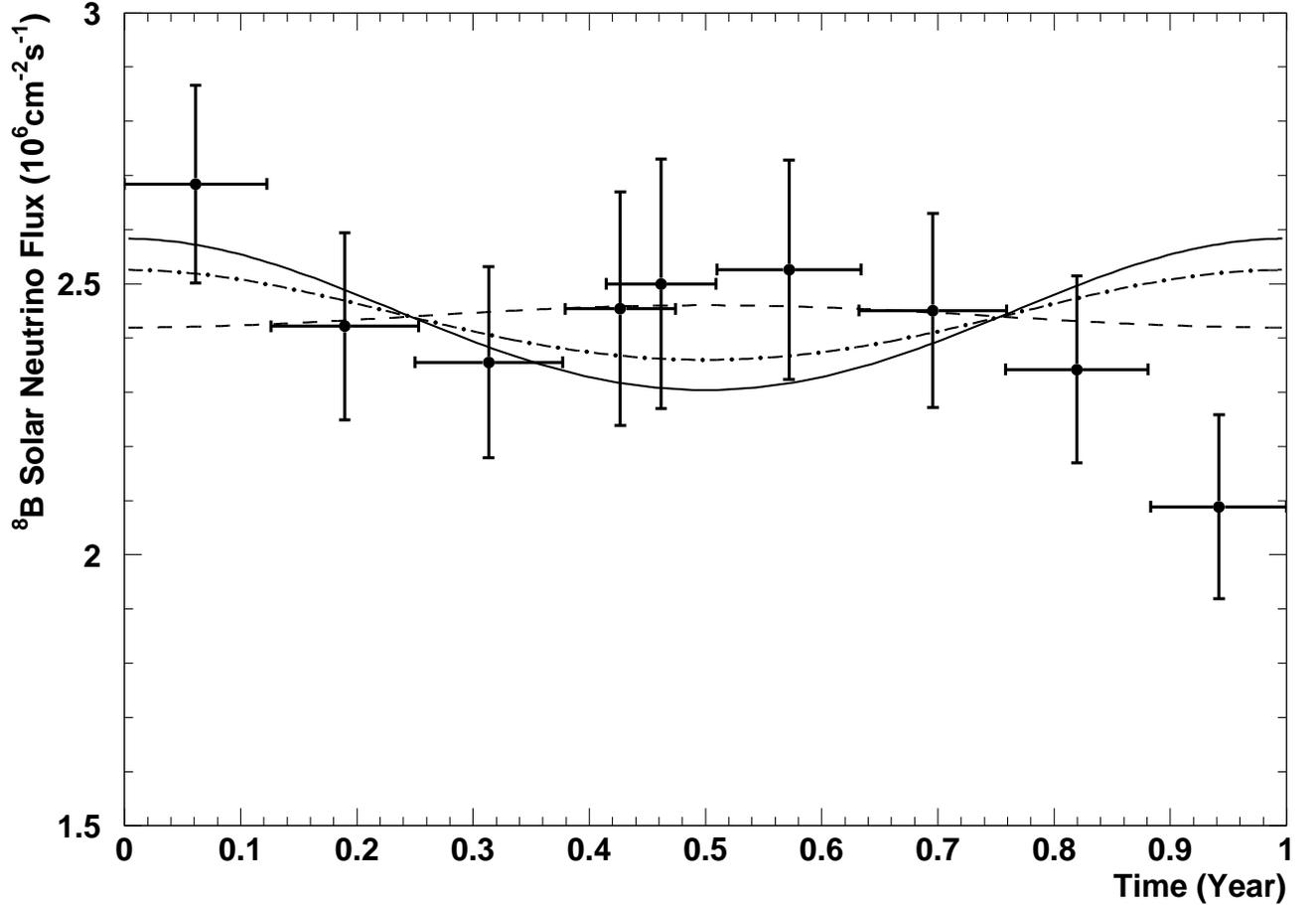,height=14.0cm}}
\protect\caption[]
{
Real time seasonal variations of the flux 
for different values of the neutrino parameters: 
$\Delta m^2 = 5.82 \cdot 10^{-11}$ eV$^2$,  
$\sin^2 2\theta = 0.90$ (solid line); 
$\Delta m^2 = 1.06 \cdot 10^{-10}$ eV$^2$,  $\sin^2 2\theta = 0.95$ 
(dashed line); 
$\Delta m^2 = 7.85 \cdot 10^{-11} {\rm eV}^2$,  $\sin^2 2\theta = 0.75$ 
(dash-dotted line). 
Dashed-dotted line coincides practically with the one 
due to pure geometrical effect.
Also shown are preliminary results from the Super-Kamiokande 
(306 days) experiment \cite{SK}. \\ 
}
\end{figure}}

\end{document}